\documentclass[journal]{IEEEtran}
\ifCLASSINFOpdf
   \usepackage[pdftex]{graphicx}
   \usepackage{amsmath}
   \usepackage{subfig}
\else
\fi
\begin{document}
%
\title{Reconstruction of Sub-Surface Velocities from Satellite Observations Using Iterative Self-Organizing Maps }
%
%
%

\author{Christopher Chapman and
        Anastase Alexandre Charantonis
\thanks{C. Chapman and is with the LOCEAN-IPSL
at the Universit\'{e} de Pierre et Marie Curie, Paris
75005 France e-mail: chris.chapman.28@gmail.com.}
\thanks{A. A. Charantonis is with SAMOVAR, T\'el\'ecom SudParis, CNRS, Universit\'e Paris-Saclay, 9 rue Charles Fourier 91011 EVRY}
\thanks{Manuscript received }}

%
%

\markboth{\tiny{This work has been submitted to the IEEE for possible publication. Copyright may be transferred without notice, 
after which this version may no longer be accessible.}}%
{Shell \MakeLowercase{\textit{et al.}}: Bare Demo of IEEEtran.cls for Journals}
%



\maketitle

\begin{abstract}
In this letter a new method based on modified self-organizing maps is presented for the reconstruction of deep ocean current velocities from surface information provided by satellites. This method takes advantage of local correlations in the data-space to improve the accuracy of the reconstructed deep velocities. Unlike previous attempts to reconstruct deep velocities from surface data, our method makes no assumptions regarding the structure of the water column, nor the underlying dynamics of the flow field. 
Using satellite observations of surface velocity, sea-surface height and sea-surface temperature, as well as observations of the deep current velocity from autonomous Argo floats to train the map, we are able to reconstruct realistic high--resolution velocity fields at a depth of 1000m. Validation reveals extremely promising results, with a speed root mean squared error of $\sim$2.8cm.$^{-1}$, a factor more than a factor of two smaller than competing methods, and direction errors consistently smaller than 30$^{\circ}$. Finally, we discuss the merits and shortcomings of this methodology and its possible future applications.       
\end{abstract}


%
\IEEEpeerreviewmaketitle

\section{Introduction}
%
%
%
%
\IEEEPARstart{S}{UBSURFACE} observations of the world's ocean, particularly of climatically interesting fields such as the velocity of ocean currents, are generally sparse both temporally and spatially. Despite recent attempts to improve ocean observing networks, our ability to directly measure oceanic properties at depth is still limited; observations obtained from ships are geographically limited, biased to warmer months in the mid-latitudes and polar oceans and are temporally discontinuous. The lack of long-term data-sets with broad spatio--temporal coverage impedes our ability to make robust inferences about changes in the climate system and limits the short-term forecasting ability of numerical models due to a dearth of subsurface data for assimilation.  

Since the early 1980s, quasi-global measurements from satellites have enabled near continuous measurement of the ocean's surface. In particular, observations of the sea-surface height anomaly (which enables direct measurement of the surface geostrophic velocity) from altimeters, and observations of the sea-surface temperature from microwave and radiometers and infrared sensors have revolutionized the understanding of the ocean's dynamics. These observations have revealed that the ocean is rich in flow features of varying spatial and temporal scales \cite{Morrow&LeTraon2012}. Despite recent efforts to expand the network of in--situ observations, the quantity and coverage of subsurface data still pales in comparison with that provided by space-borne instruments\cite{RiserEtAl2016}.

\begin{figure*}[!t]
\centering
\subfloat{\includegraphics[width=3.0in,height=3.6in]{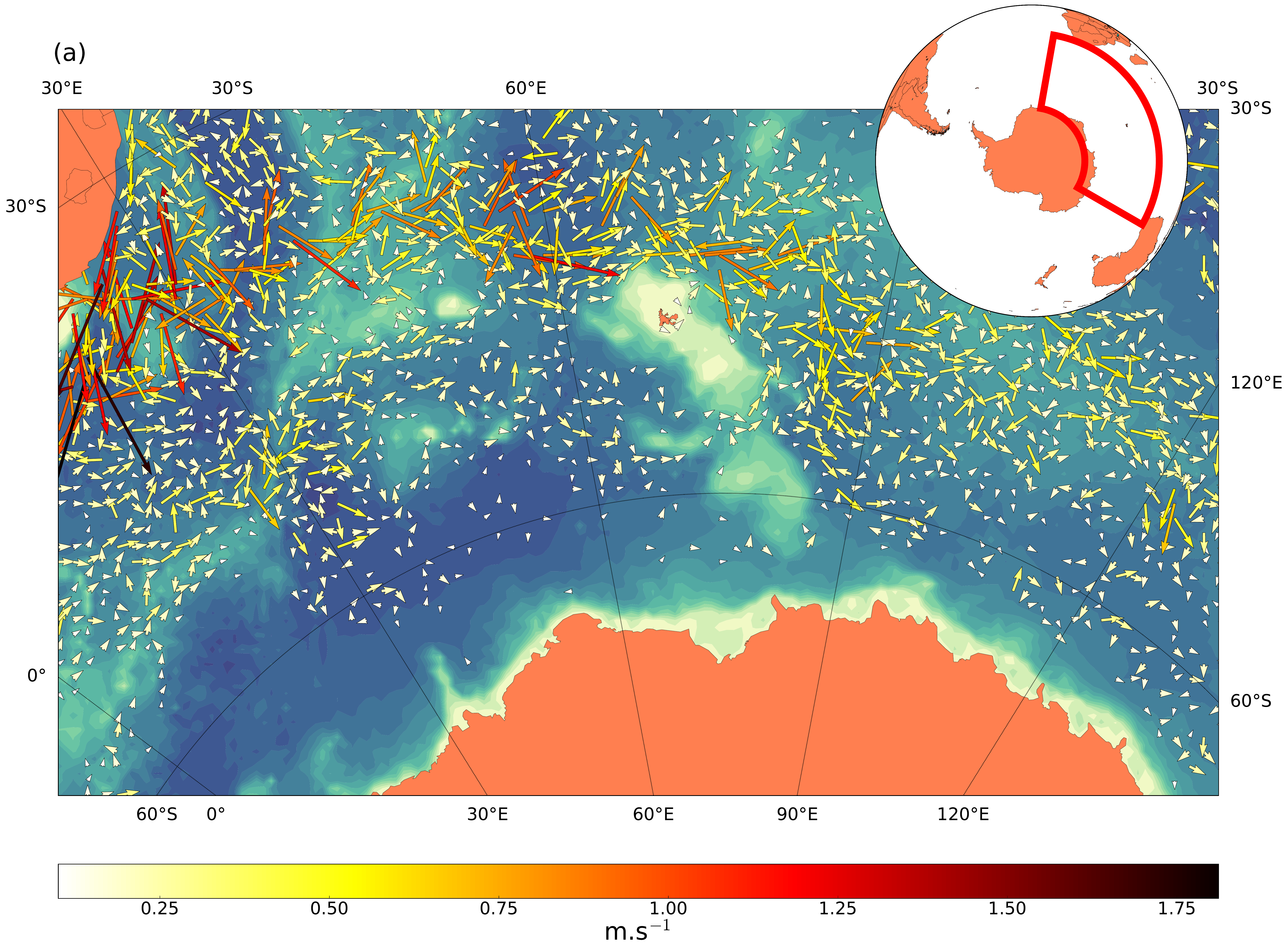}%
\label{fig_first_case}}
\hfil
\subfloat{\includegraphics[width=3.0in,height=3.4in]{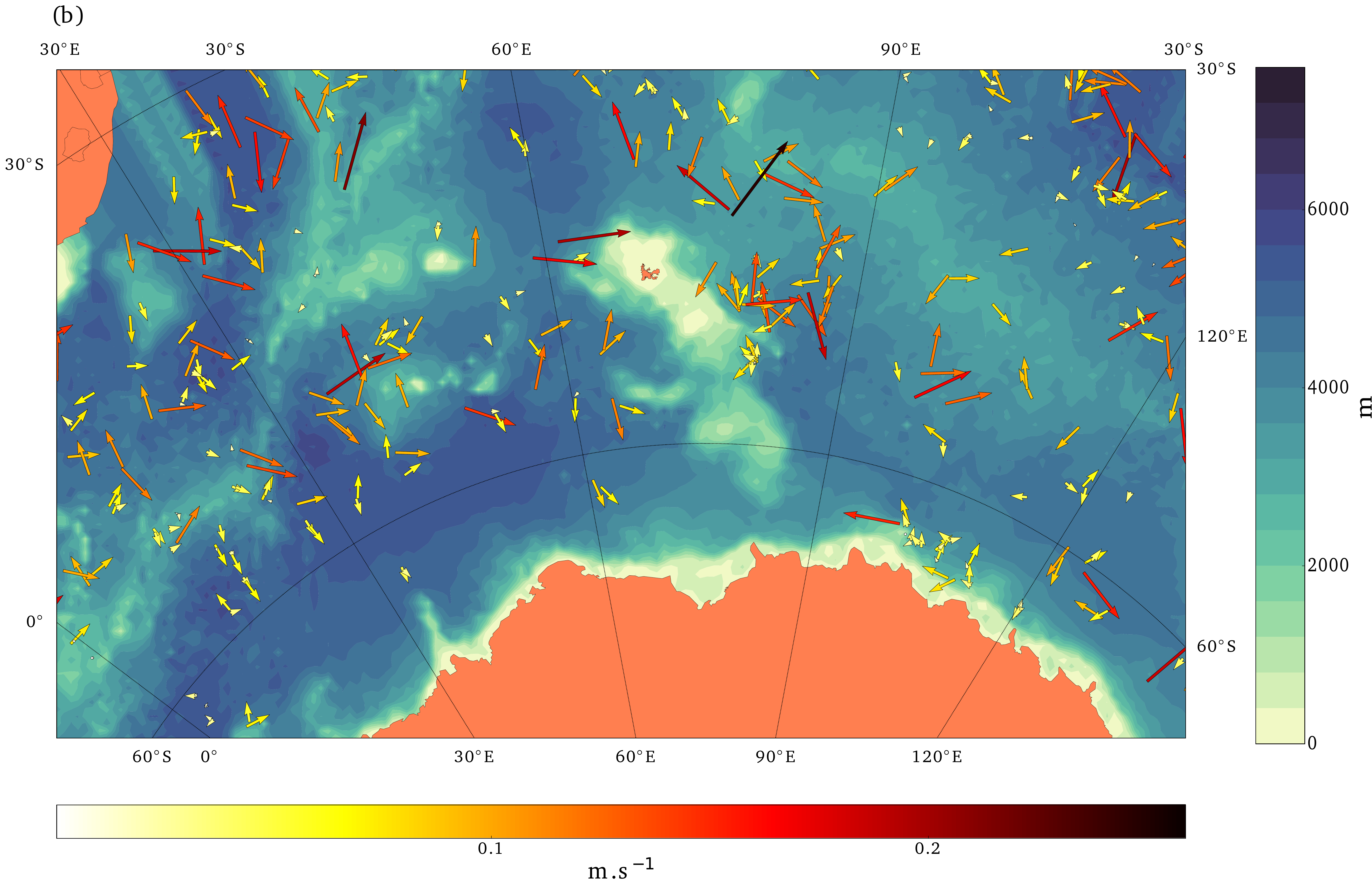}%
\label{fig_second_case}}
\caption{Current velocity at the (a) surface from the AVISO satellite produce; and (b)  near 1000m depth from the Argo drifters in the South Indian Ocean, near the Kerguelen Plateau, on the 17th of April, 2009. The location of this region is indicated in the inset box. The right-hand side color--bar indicates the ocean depth. Note the differing color scales between each panel.}
\label{Fig:Intro_Data}
\end{figure*}
We illustrate the difference in the spatial coverage between surface and deep measurements in Fig. \ref{Fig:Intro_Data}, which shows the surface velocity (Fig. \ref{Fig:Intro_Data}(a)) from satellite altimetry  and at approximately 1000m depth from Argo floats (Fig. \ref{Fig:Intro_Data}(b)) in the South Indian Ocean on the 17th of April, 2009. It is clear that the satellite data provide broad spatial coverage of the region, while the measurements at depth are scattered, often with large distances between measurements. This gap in coverage has lead to numerous efforts to reconstruct sub-surface quantities from high-resolution satellite data. 

Traditionally, attempts to reconstruct the deep flow from surface observations fall into two different categories that we label ``statistical'' or ``dynamical'' methods, although on closer-inspection both methodologies make similar assumptions about the underlying state of the ocean. ``Statistical'' methods take advantage of empirical relationships between surface and subsurface quantities to reconstruct the subsurface fields, subject to the assumptions that these relationships are static in time and that the vertical structure of the water column can be represented as a simple function of depth \cite{Sun&Watts2001,ParkElAt2005,Willis2010,MeijersEtAl2011}. In contrast ``dynamical'' methodologies combine the equations of fluid motion with surface information from satellites to estimate the sub-surface fields \cite{IsernFontenetEtAl2008,Keating&Smith2015}. However, dynamical reconstructions require the basic stratification of the ocean to be slowly varying in space and predict the sub-surface fields by inverting an elliptic partial differential equation, which simply smooths and attenuates the surface fields. Thus, the reconstructed fields smooth out small scale structures.

Recently, machine learning techniques have been applied to similar problems \cite{Liu&Weisberg2011}. Machine-learning methods have several advantages over the methods described above: the relationships between surface and sub-surface quantities can vary in space and time; it is not necessary to make any assumptions about the vertical structure of the water column; and non-linear relationships can be extracted from the data in a non-supervised way. Example applications include the use of self-organizing maps (SOMs) reconstruct chlorophyll profiles from ocean color images \cite{CharantonisEtAl2015}; or for the completion of a database of hydrographic profiles obtained from ocean gliders\cite{Charantonis2015a}. 

In this letter, we tackle the problem of reconstructing sub-surface velocities from surface data using a methodology based on self-organizing maps. We restrict our attention to the Southern Ocean, the region of the world that encircles the Antarctic continent and lies south of Australia, South America and Africa. We focus on this region as it hosts an energetic and complex flow field that presents a challenge for prediction schemes, and due to its remote location and harsh environment, is one of most data sparse ocean basins. Hence, a robust reconstruction of the deep flow from satellite data could be of immense benefit to the oceanographic community.  

The remainder of this paper is organized as follows: in section \ref{Section:Data_and_Methods} we introduce our methodology, based on the method of Charontonis \textit{et al.} \cite{Charantonis2015a} for completing data sets with missing or corrupted data , and the data we will use is described. We validate our methodology and present examples of the reconstructed deep flow in section \ref{Section:Results}. Finally, in section \ref{Section:DiscussionConclusion}, we discuss the potential applications of this work, as well as its shortcomings and avenues for future improvements.  

\section{Data and Methodology} \label{Section:Data_and_Methods}

\subsection{Data}

The data used in this study consists of \textit{surface data} that are used as predictors, and \textit{sub-surface} measurements of current speed that are used to train the SOM and for validation.

\subsubsection{Surface Data}

In order to reconstruct the sub-surface current velocities, we use as predictor variables the surface velocity, surface absolute dynamic topography and the sea surface temperature obtained from satellite data. 

The dynamic sea-surface topography used in this study is obtained from the Archiving, Validation, and Interpretation of Satellite Oceanographic data (AVISO) weekly gridded sea level \textit{anomalies} (SLA) (http://www.aviso.altimetry.fr/), a ``level 4" product. We use SLAs for the 5 year period 2005--2011, mapped to a 1/4 degree Mercator grid using optimal interpolation of along-track data series using at least two satellite missions [TOPEX/Poseidon--ERS or Jason-1--Envisat or Jason-2--Envisat] with consistent sampling over the time period. This dataset provides estimates of the sea-surface height and velocity \textit{anomalies}, relative to the 20 year period from 1994 to 2014. The time mean sea-surface height and current velocity are obtained using the mean dynamic topography, reconstructed by combining data from the Gravity Recovery and Climate Experiment (GRACE) mission, satellite altimeters, and drifting buoys \cite{Rio2014a}. An example of the surface velocity  from these datasets is shown in \ref{Fig:Intro_Data}(a).  

The sea--surface temperature (SST) data used are daily averages of Version 2 of the NOAA combined AVHRR-AMSR optimally interpolated SST product (https://www.ncdc.noaa.gov/oisst)\cite{ReynoldsEtAl2007}. The combined use of infrared and microwave instruments in cloud-free regions reduces systematic biases as the errors of each sensor are independent. 

\subsubsection{Sub-Surface Velocity Data} 
In order to estimate the ocean current velocity at depth, we use the velocity data provided by autonomous lagrangian drifters called Argo floats \cite{RiserEtAl2016}. After deployment, Argo floats descend to a pre--programmed ``parking" depth (generally 1000m) where they drift with the current for approximately 10 days, then ascend to the surface (taking a profile of temperature and salinity) and transmit their location by satellite to a data center. The floats then re-descend to their parking depth and repeat the cycle. With knowledge of the time between each surfacing, as well as the distance between the surfacing locations, one can estimate the parking depth velocity \cite{Ollitrault&Rannou2013}. \\ 
In this study, we make use of the ANDRO data set (http://www.umr-lops.fr/Donnees/ANDRO), described by Ollitrau \& Rannou \cite{Ollitrault&Rannou2013}. This dataset provides estimates of the current velocity at the float parking depth between 2005 and 2011 and covers the entire Southern Ocean north of about 65$^{\circ}$S, although, as can be seen in Fig \ref{Fig:Intro_Data}(b), the data are not evenly spatially distributed. There are 122,174 independent data records in the dataset and errors due to the delay between the float surfacing and the satellite location fix and vertical shear in the water column are estimated to be small.  

In this study, we restrict our attention to the Southern Ocean between 65$^{\circ}$S and 35$^{\circ}$S. We also use only floats with parking depths within 50m of 1000m, as there are sufficient floats at this depth to enable broad coverage of the Southern Ocean. In contrast, the number of floats at parking depths different from 1000m is much more limited and they are not able to provide broad geographic sampling\cite{Ollitrault&Rannou2013}.

\subsection{Methodology}

\begin{figure}[!t]
\centering
\includegraphics[width=3.75in]{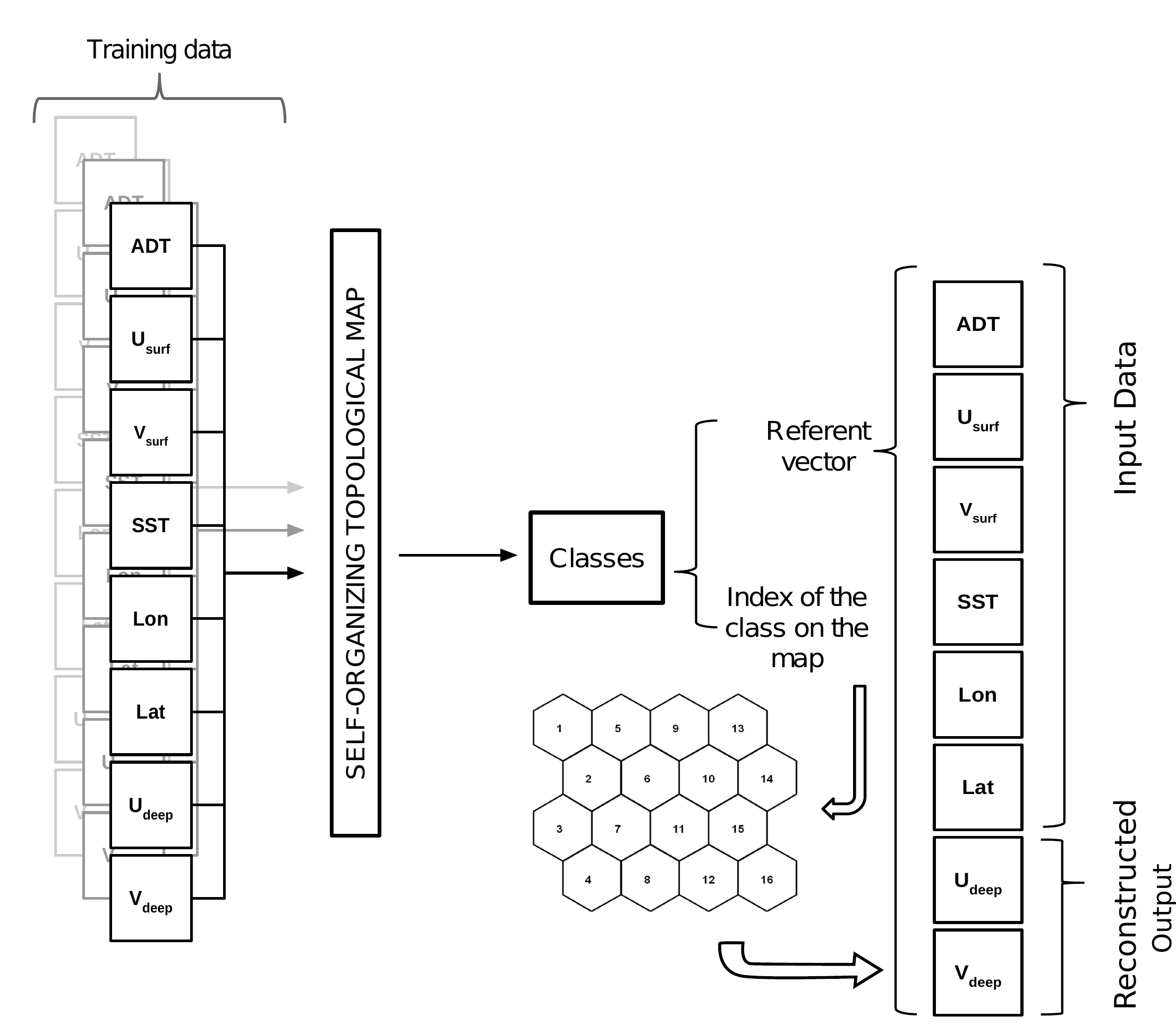}
\caption{Schematic of our methodology. On the left, we show the data used to train the SOM, which maps the training data to a discreet set of classes. Each class contains a referent vector containing, for each parameter, the average value of the elements comprising it as well as an index of the class that informs us of its location on the topological map. The right side shows the reconstruction: the available (surface) components of the referent vector are projected via a similarity function onto the SOM and the missing (deep) values are extracted from the best-matching unit.}
\label{fig:Method_Diagram}
\end{figure}

Self-Organizing Maps are a neuronal network classification algorithm incorporating a topological structure used for organization of the different classes on a 2D lattice. Each class is represented by a referent vector in the data space and an index positioning it on the 2D lattice. The referent vectors of two neighboring classes on the 2D lattice are by construction close in the data space. After the initial training, the SOM map can be used to predict missing data, shown schematically in Fig. \ref{fig:Method_Diagram}. This is generally done by projecting the available values (i.e. at the surface) on the SOM and completing the missing values (i.e. the values at depth) with the corresponding values of the best-matching unit, which corresponds to the referent vector that is closest in the euclidean sense to the input vector.

This approach is, however, not optimal. Due to local correlations in the data space, the pertinence of each parameter to the retrieval of a missing parameter's value varies throughout the data space. As such, we used a similarity function (defined in Charantonis \textit{et al}. $\cite{Charantonis2015a}$) when performing this comparison. This similarity function between a vector $X$ in the data space containing missing and non-missing components and a referent vector $\textit{ref}^{c}$ of the SOM class $\textit{c}$, denoted $\textbf{\textit{s}}^{\textit{c}}( \textit{X}, \textit{ref}^{c})$ is defined as:
\begin{equation} \label{Eqn:Local_Corr}
\begin{split}
\textbf{\textit{s}}^{\textit{c}}( X, \textit{ref}^{c}) = \sum_{i \in non-missing} \left( 1+ \sum_{j \in missing}\left(\textit{cor}^{ c}_{i,j}\right)^{2} \right)\dots \\ \times \sqrt{(X_i-ref^c_i)^2},
\end{split}
\end{equation}
where $\textit{cor}^{ c}_{i,j}$ is the local correlation between the missing and non-missing variables computed over the data attributed to the class $\textit{c}$ during the training phase of the SOM algorithm. In the case were there are insufficient data points to calculate this local correlation, we exploit the topology of the SOM by using the data that belong to neighboring classes (and therefore are close in the data space) in order to calculate this local correlation.

\section{Reconstruction of Deep Velocities from Satellite Data} \label{Section:Results}

\subsection{Validation and Errors}
The SOM methodology is now applied to the problem of reconstructing velocity fields at depth from the satellite observations described in section \ref{Section:Data_and_Methods}. As inputs we use the surface velocity, dynamic height and temperature from satellite observations, as well as the deep velocity obtained from the Argo floats and the latitude and longitude of each deep observation. 
 Surface data is co--located at the subsurface data locations by linear interpolation.  80\% of this data-set, (97,739 data records), is selected by random sampling and used to train the SOM. The remaining 20\% (24,435 records) are retained for validation. 

\begin{figure}[!t]
\centering
\includegraphics[width=3.5in]{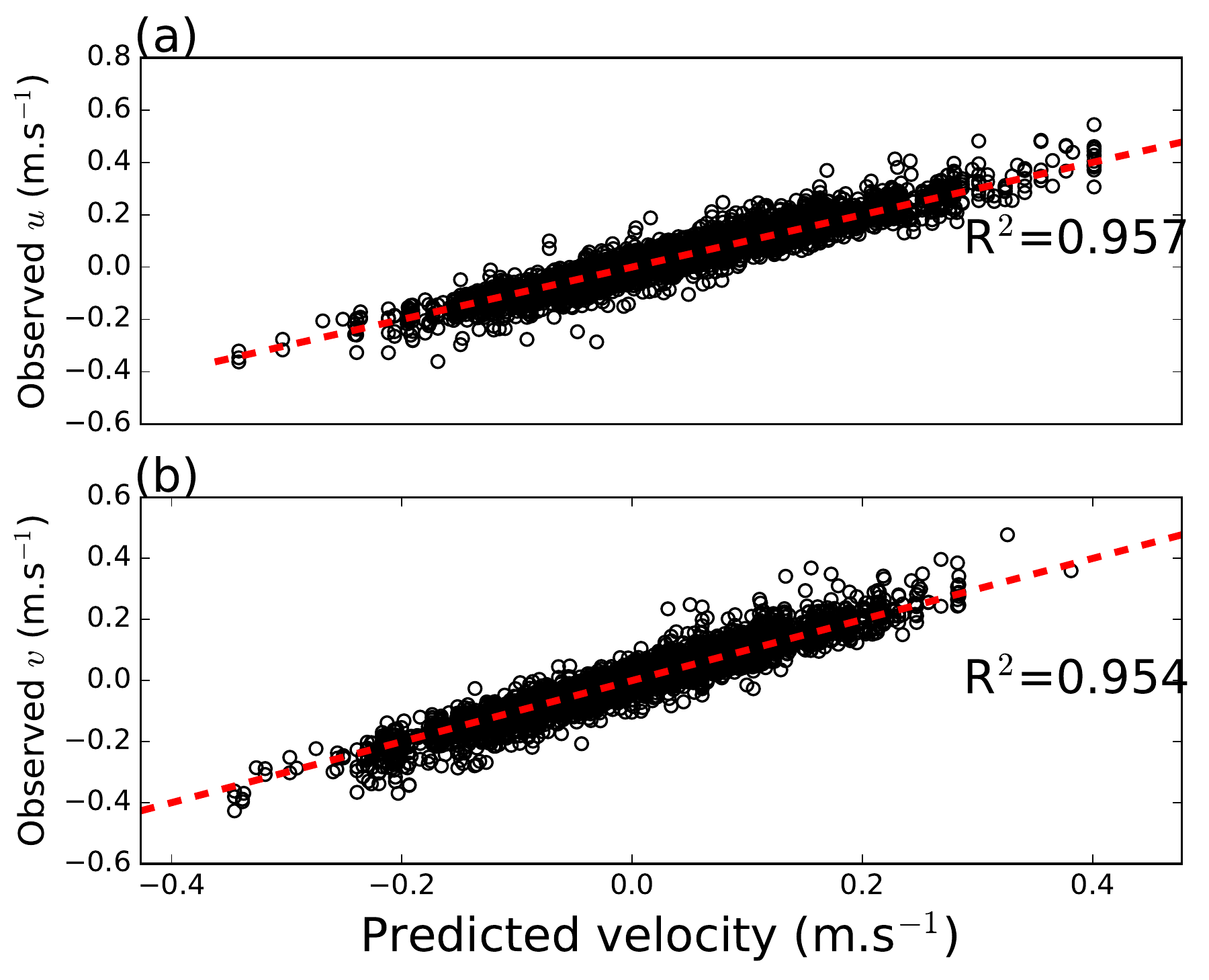}
\caption{The reconstructed versus observed zonal (a) and meridional (b) velocities over the Southern Ocean at 1000m. The red dashed line indicates the ``perfect'' reconstruction. $R^2$ values for each series are indicated.}
\label{fig:Error_Scatter_Plots}
\end{figure}

The validation of the method, using a map trained with 3000 data classes is shown in Fig \ref{fig:Error_Scatter_Plots}. It is evident from this figure that our results are very promising. The $R^2$ of each velocity component is above 0.95, and the speed RMSE is 2.8cm.s$^{-1}$, more than a factor of 2 smaller than those obtained by Meijers \textit{et al.} \cite{MeijersEtAl2011} and a factor of 3 smaller than those obtained from dynamical type methods such as Isern--Fontenet \textit{et al} \cite{IsernFontenetEtAl2008}. To ensure that our method is not subject to over--fitting, we also validate against the training data, obtaining a speed RMSE of 2.6cm.s$^{-1}$. This value is sufficiently close to the RMSE obtained from the validation data that we can rule out over--fitting. We note that using the SOM methodology on its own, without the correction described by Eqn. \ref{Eqn:Local_Corr}, gives inferior results, with RMSEs of $\sim$7.5cm.s$^{-1}$ and $R^{2}=\sim$0.6.  

\begin{figure}[!b]
\centering
\includegraphics[width=3.9in]{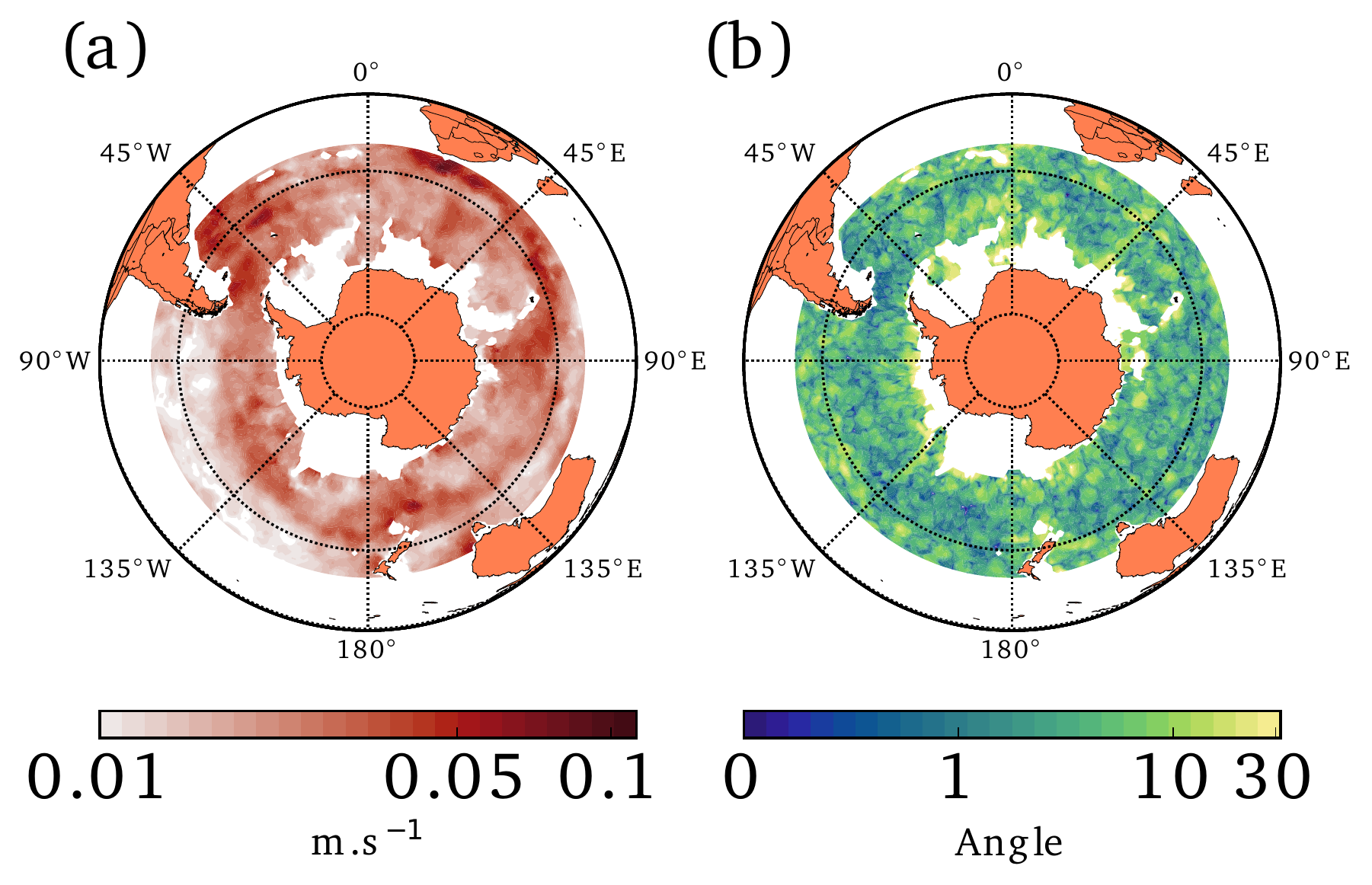}
\caption{Maps of the mean error in the reconstructed (a) speed; and (b) bearing angle.  Note the logarithmic colorscale.}
\label{fig:Error_Maps}
\end{figure}

We compute the spatial distribution of the errors in both the current speed and direction, quantified by the bearing angle $\theta=\tan^{-1}\left(v/u \right)$, by determining the error at the location of each deep velocity observation in the validation dataset, then mapping the results to a regular latitude/longitude grid with 0.5$^{\circ}$ by taking the ensemble mean of the errors  within 150km of every point on that grid. The results of this calculation are shown in Fig \ref{fig:Error_Maps}. 

Current speed errors, $\epsilon_\textrm{speed}$, (Fig. \ref{fig:Error_Maps}(a)) are concentrated in certain geographic regions. Similarly to Meijers \textit{et al} \cite{MeijersEtAl2011}, we find elevated $\epsilon_\textrm{speed}$ downstream of large sub--surface topography, (e.g. downstream of the Kerguelen Plateau at $\sim$80$^{\circ}$E). Additionally, $\epsilon_\textrm{speed}$ correlates with the SLA variance, a proxy for ocean meso--scale turbulence, which is known to be enhanced downstream of large sub-surface topographic features \cite{WilliamsEtAl2007,ChapmanEtAl2015}. The increasing error in highly turbulent regions suggests a decoupling of the surface and deep flow that may limit the effectiveness of  reconstructions based on surface data. In contrast, errors in the bearing angle show no geographic concentration and appear to be distributed quasi--randomly. We note that more than 85\% of bearing--angle errors are less that 30$^{\circ}$, with a mean absolute error of 18$^{\circ}$, a median of 10$^{\circ}$ and no identifiable bias--that is bearing--angle errors are distributed symmetrically around 0. 

\subsection{Reconstruction of Deep Southern Ocean Currents}

We now use our method to reconstruct maps of deep currents over the entirety of the Southern Ocean basin. To do this, we apply our method to each of the daily output maps in the AVISO and OISST databases, between 2005 and 2011. We obtain 5 years of daily current velocities (1826 snapshots) at 1000m on a regular latitude/longitude grid with 1/4$^{\circ}$ grid spacing. Grid points with fewer than 10 deep velocity observations within 150km (typically south of 65$^{\circ}$S) are masked.    

\begin{figure}[!t]
\centering
\includegraphics[width=3.75in]{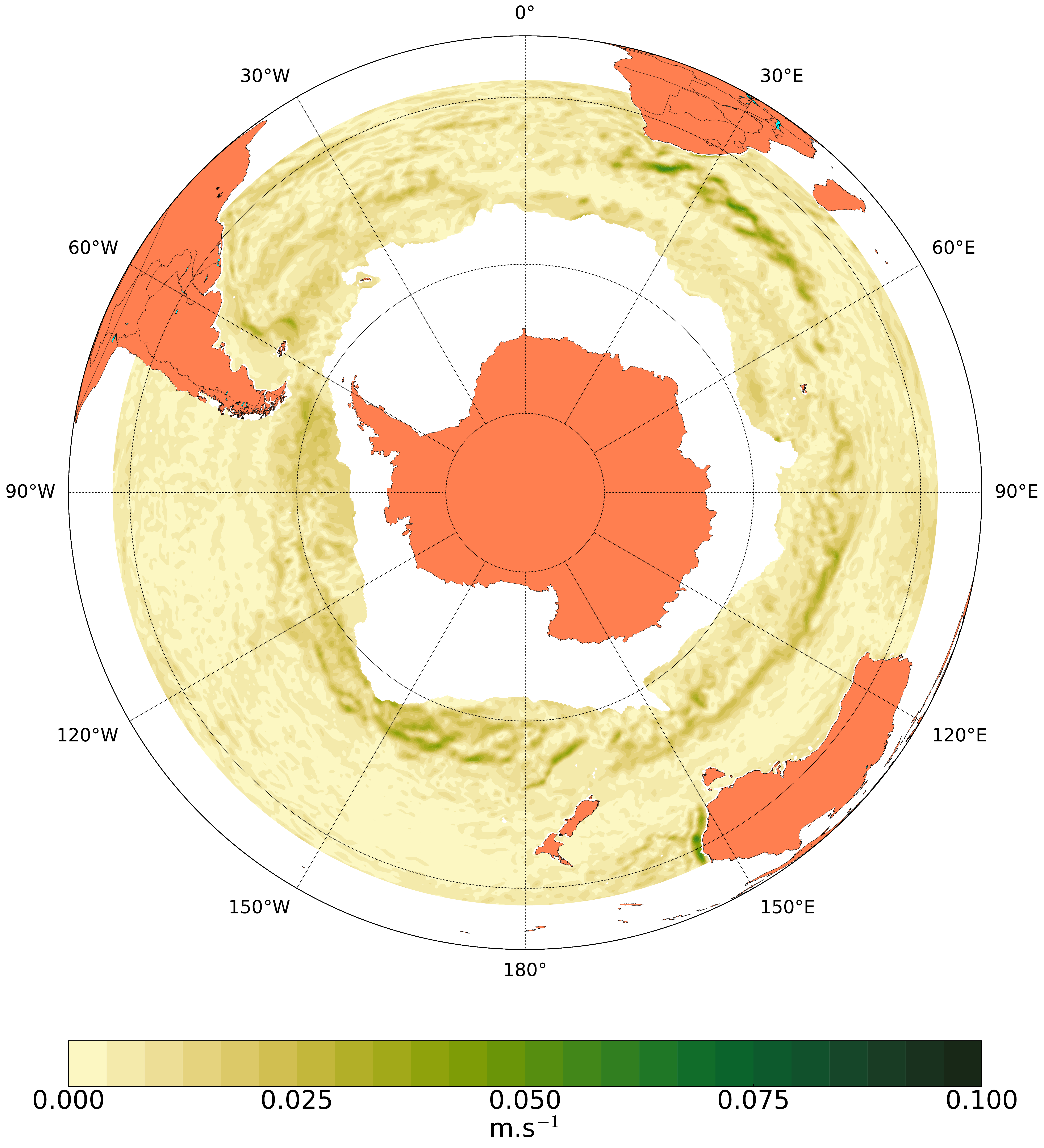}
\caption{The time mean reconstructed current speed at 1000m depth for the year 2009.}
\label{fig:Reconstructed_Currents_2009}
\end{figure}

Fig. \ref{fig:Reconstructed_Currents_2009} shows the reconstructed time mean current speed for the year 2009. Our reconstructed currents display realistic behavior such as self-organisation into complicated small--scale ($\mathcal{O}$(10--20km)) ``jets'' (e.g. south of Africa between 20$^{\circ}$E and 50$^{\circ}$E), steering by subsurface topography (e.g. south of New Zealand between 170$^{\circ}$E and 170$^{\circ}$W) and western boundary currents (e.g. along the east coast of Australia and South America). Animations of the reconstruction additionally reveal realistic propagation of meso-scale eddies and Rossby waves.  

Our reconstructed deep velocity maps show similar features to the satGEM reconstruction of Meijers \textit{et al}. \cite{MeijersEtAl2011}. However, there are some notable qualitative differences between the two reconstructions. Firstly, we find that our reconstructed currents are generally slower than the satGEM product, locally by as much 20\%. Secondly, we note that our flow fields are also smoother than the satGEM flow fields. This smoothness arises due to the finite number of classes used in our SOM. Inputs that are in the same neighborhood in data-space (thus close in geographical space) tend to be sorted into the same class and therefore yield an identical reconstructed result.        

\section{Discussion and Conclusions} \label{Section:DiscussionConclusion}

In this letter, we have used a machine-learning technique to reconstruct the velocity of ocean currents at 1000m depth from satellite observations. Our results yield errors  that are 2 to 3 times smaller than competing methods. We are able to use this method to reconstruct realistic maps of deep currents with high temporal and spatial resolution. 

Despite the promising results, our methodology has several shortcomings. Most notably, to train the SOM we require velocity information at depth. While we have been able to exploit the near global coverage provided by Argo floats at 1000m, velocity data are more limited at other depths \cite{Ollitrault&Rannou2013}, which reduces our ability to apply this method more generally. The satGEM dataset of Meijers \textit{et al.} \cite{MeijersEtAl2011} and dynamical methods \cite{IsernFontenetEtAl2008} are not limited by the availability of deep velocity data and can provide reconstructions at any depth.  

Despite this shortcoming, our method has numerous potential applications beyond the obvious extension to other quantities, such as temperature and salinity profiles. Due to its relatively modest computational expense, the SOM method could be used for real-time for data-assimilation into predictive ocean models, or for validating numerical models in data-sparse regions. Additionally, although we have made use of gridded ``level 4'' satellite products in this letter, the method could also make use of ``level 3" along-track products without substantial modification for real--time application.

\appendices
\section{Data and Code Availability}
Code, written in MATLAB, is available from the author's GitHub site: 
https://github.com/ChrisC28/SOM\_Reconstruction
NetCDF files containing the reconstructed velocities at 1000m from 2005 to 2011 can be downloaded from:https://www.dropbox.com/sh/vltobmd3cz7d7eq/AAAyxp8qoeW9ArWRyUyvk1-3a?dl=0

\section*{Acknowledgment}
C.C. is supported by a National Science Foundation Ocean Sciences Directorate Postdoctoral Fellowship, \#1521508.
This work is supported by the Center for Data Science, funded by the IDEX Paris-Saclay, ANR-11-IDEX-0003-02.

\ifCLASSOPTIONcaptionsoff
  \newpage
\fi

\end{document}